# Origin of dielectric relaxation observed in complex perovskite oxides $Ba_{1-x}La_xTi_{1-x}Cr_xO_3$

Mamoru Fukunaga[a)] and Yoshiaki Uesu

Department of Physics, Waseda University, Tokyo 169-8555, Japan

Frequency dependence of the dielectric constant of complex perovskite oxide $Ba_{1-x}La_xTi_{1-x}Cr_xO_3$ (BLTC) ceramics with the composition ratio $0.4 \leq x \leq 0.7$ is precisely measured in the temperature range from 20 K to 300 K, and the dielectric relaxation is found to be quite similar to that of $CaCu_3Ti_4O_{12}$ (CCTO), which exhibits the Debye-like frequency dispersion around 100 K. In BLTC, the ferroelectric phase transition temperature shifts to lower temperature with increase of $x$, while a remarkable dielectric relaxation newly appears at higher temperature region. The dielectric relaxation can be explained by an equivalent model of a series of two $R$-$C$ parallel circuits, which corresponds to a heterogeneous structure in the sample with an internal barrier layer capacitor. Temperature and frequency dependences of the measured dielectric constant are explained well by the model with the temperature-dependent electrical conductivity and temperature-independent intrinsic dielectric constant. It is also found that the contribution of thin layer to the dielectric relaxation in a high temperature region is affected by the dc bias field on the sample.

---

[a)] Electronic mail: fukunaga@ruri.waseda.jp

# I. INTRODUCTION

$Ba_{1-x}La_xTi_{1-x}Cr_xO_3$ (BLTC) ceramics form perovskite type solid solutions in the entire range of the ratio $x$ from $x = 0$ ($BaTiO_3$) to $x = 1$ ($LaCrO_3$), and the crystal structure at room temperature changes succesively from tetragonal to cubic, rhombohedral, and orthorhombic as $x$ increases with intermediate two-phase coexisting region (Fig. 1).[1] The lattice constants of BLTC change linearly with $x$ in each phase except for two-phase coexisting region, and the dielectric properties also change continuously with $x$ as shown in Fig. 2 for $x = 0.04, 0.06, 0.08$ and $0.10$. The peak of dielectric constant $\varepsilon'$ (real part) corresponds to a ferroelectric phase transition, and the transition temperature decreases by approximately 37 K per 1 % of $x$.[2] We have recently found that samples with $x \geq 0.08$ behave like relaxors[3], and these are candidates of lead-free relaxors. The feature of temperature changes of $\varepsilon'$ and $\varepsilon''$ (imaginary part) for each frequency seem to shift to lower temperature as $x$ increases. The increase of $\varepsilon''$ with increasing temperature and the inverse proportion to the frequency indicate the existence of the dc electric conductivity.

Recently, another complex oxide $CaCu_3Ti_4O_{12}$ (CCTO) has attracted much attention because of very high and flat temperature dependence of dielectric constant around room temperature.[4] It arises also interest on the sudden depression of the dielectric constant with the Debye-like dispersion below about 100 K without any structural phase transition.[5] Whether the origin of the dielectric behavior of CCTO is intrinsic or extrinsic has been discussed so far but the problem is still open.[6]

This paper reports that the dielectric relaxation observed in BLTC ceramics with $0.4 \leq x \leq 0.7$ below room temperature is similar to CCTO, and can be explained by a simple heterogeneous structure model of equivalent $R$-$C$ circuits. Similar experimental results and consideration have already been given for CCTO[7], $AFe_{1/2}B_{1/2}O_3$ (A = Ba, Sr, Ca; B = Nb, Ta, Sb)[8], $Gd_{0.6}Y_{0.4}BaCo_2O_{5.5}$[9], and $La_{0.5}Sr_{0.5}Ga_{0.6}Ti_{0.4}O_{3-\delta}$[10] ceramics, and In-doped $CdF_2$[9], $LaMnO_3$[9], $SrNbO_{3.41}$[9], and hexagonal $BaTiO_3$[11] single crystals, and so on. As far as we know, however, this paper is the first example of well-known perovskite $BaTiO_3$ based ceramics. In addition, we suggest that the measurement of the dc bias effects can provide more precise information on the heterogeneous structure.

## II. EXPERIMENTAL

BLTC ceramic samples were synthesized from stoichiometric amounts of $BaCO_3$, $TiO_2$, $La_2O_3$, and $Cr_2O_3$. The samples were sintered at 1500°C in air. Details of the sample synthesis including the difference of sintering temperature are given in Ref. 1. Typical dimensions of disk-shaped samples were 12.6 mm in diameter and 2.0 mm in thickness. Conductive silver paste (Fujikura-Kasei DOTITE D-500) was painted on the circular faces as electrodes and burnt at 400°C.

The complex impedance of the samples was measured by a Solartron SI-1255B frequency response analyzer with an SI-1296 dielectric interface. $\varepsilon'$ and $\varepsilon''$ were calculated from the impedance and sample dimensions. Applied ac amplitude was 1 $V_{rms}$. Sample temperature was changed from room temperature to 20 K with frequencies from 100 kHz to 1 Hz. Measurements under dc bias voltage up to 40 V were performed at 30°C with frequencies from 1 MHz to 10 mHz.

## III. EXPERIMENTAL RESULTS

Figure 3 shows temperature dependence of the dielectric constant of the BLTC sample of $x = 0.50$ (1.95 mm thick, hereafter termed BLTC-0.50). $\varepsilon'$ is over 4000 at 295 K and decreases to about 30 towards 50 K with remarkable frequency dispersion. $\varepsilon''$ shows a peak at the center temperature of the dispersion for each frequency. Temperature and frequency dependences of both $\varepsilon'$ and $\varepsilon''$ roughly follow the Debye-type dielectric relaxation described as,

$$\varepsilon'(\omega) = \varepsilon(\infty) + \frac{\Delta\varepsilon}{1+(\omega\tau)^2},$$
$$\varepsilon''(\omega) = \frac{\Delta\varepsilon\omega\tau}{1+(\omega\tau)^2}, \tag{1}$$

where

$$\Delta\varepsilon = \varepsilon(0) - \varepsilon(\infty),$$
$$\tau(T) = \tau_0 \exp(E/kT).$$

High and low frequency limits of the dielectric constant, $\varepsilon(\infty)$ and $\varepsilon(0)$ correspond to $\varepsilon'$ at low and high temperature, respectively. The relaxation time $\tau_0$ and its activation energy $E$ can be obtained by the Arrhenius plot of the peak temperature for $\varepsilon''$ and the frequency $f = \omega/2\pi = 1/2\pi\tau$. The values $\varepsilon(\infty) = 30$, $\varepsilon(0) = 4000$, $\tau_0 = 2.29$ ns, and $E = 153$ meV are obtained for BLTC-0.50. The values for other BLTC samples with different $x$ are listed in Table 1. Both $\varepsilon(\infty)$ and $E$ tend to decrease as $x$ increases, but dependence of $\varepsilon(0)$ and $\tau_0$ on $x$ is not clear. The crystal structure of BLTC is cubic for $x = 0.40, 0.50$ and $0.60$, and rhombohedral for $x = 0.70$, however, the difference in the dielectric properties seems not to be directly related to the structure.

Figure 4 shows the relation between real ($Z'$) and imaginary ($Z''$) parts of the impedance under different dc bias voltages for (a) BLTC-0.50 and (b) the sample of $x =$ 0.03 (2.13 mm thick, BLTC-0.03) at 30°C. Both samples show almost the same $\varepsilon'$ about 4700 at 10 kHz, however, the ferroelectric phase transition temperature of the BLTC-0.03 is 29°C and the sample exhibits high $\varepsilon'$. Semicircles of BLTC-0.50 diminish by the dc bias as shown in Fig. 4(a), which means the decrease of the sample resistance. On the other hand, the BLTC-0.03 shows much larger resistance than that of the BLTC-0.50 and no change under the dc bias as shown in Fig. 4(b). Figure 4(c) shows the comparison of $Z'$-$Z''$ plots of BLTC-0.50 and BLTC-0.03 in high frequencies. It is disclosed that the resistance about 160 Ω exists in series to the resistance corresponding to Fig. 4(a) for the BLTC-0.50, but such a resistance does not exist for the BLTC-0.03. The dc bias effect was not observed for both samples in the area as Fig. 4(c).

## IV. DISCUSSION

From the results shown in Fig. 4(a) and (c), an equivalent circuit of a series of two $R$-$C$ parallel connections (Fig. 5(a)) is considered for the BLTC-0.50. In this case, $R_1$ and $R_2$ correspond to $Z'$ at the left edge (magnified in Fig. 4(c)) and the width of the semicircle in Fig. 4(a), respectively.[12] The dc bias affects only $R_2$. The capacitance $C_1$ and $C_2$ are related to the frequency $f_n = (2\pi R_n C_n)^{-1}$ ($n = 1, 2$) which determines the top of the complex impedance semicircles. From the experimentally determined values of $R_2 = 11.6$ MΩ and $f_2 = 1.5$ Hz, $C_2$ is calculated to be 9.2 nF for BLTC-0.50 at 30°C without dc bias. $C_1$ is estimated to be much smaller than 1 nF, because $R_1 = 160$ Ω and the absence of complex impedance semicircle in this experiment means $f_1 \gg 1$ MHz.

When $R_1 \ll R_2$ and $C_1 \ll C_2$ ($f_1 \gg f_2$) as in BLTC-0.50 and these are frequency-independent, the resultant complex capacitance $C' - iC''$ of the circuit in Fig. 5(a), which is proportional to $\varepsilon'$ and $\varepsilon''$ of the sample, can roughly be explained as follows. At a high frequency $f \gg f_1$, an impedance $Z = (i\omega C)^{-1}$ of a capacitor $C$ becomes smaller than that of a resistor, and then $R_1$ and $R_2$ can be ignored as Fig. 5(b). Then, the resultant capacitance is close to $C_1$. When $f \ll f_2$, the resultant impedance of $R_1$ and $C_1$ can be ignored in comparison with that of $R_2$ and $C_2$ as in Fig. 5(d), then $C_2$ and $R_2$ appear as a large capacitance with a loss causing the increase of $\varepsilon''$. If $f$ is in the intermidiate range $f_1 \gg f \gg f_2$, $C_1$ and $R_2$ are ignored as in Fig. 5(c), then $C'$ and $C''$ are expressed as Eq. (2). The expressions are very similar to Eq. (1), and the measured dielectric constant shows the Debye-like frequency dispersion.

$$C'(\omega) = \frac{C_2}{1+(\omega\tau)^2},$$
$$C''(\omega) = \frac{C_2 \omega\tau}{1+(\omega\tau)^2}, \qquad (2)$$
$$\tau = R_1 C_2.$$

Resistance $R$ of dielectric (or semiconductive) materials generally decreases as temperature increases, thus the high frequency case (Fig. 5(b)) corresponds to the low temperature one, while the low frequency case (Fig. 5(d)) to the high temperature one. In the specific temperature and frequency range, the condition for Fig. 5(c) is fulfilled and an apparent (unrelated to the polarization) dielectric relaxation appears.

Temperature dependence of $R$ of dielectric materials can usually be written as $R = R_0 \exp(E/kT)$ with the activation energy $E$. Therefore if $C$ is independent of temperature, Eq. (2) almost agrees with Eq. (1). In other words, the temperature dependence of $R_1$ causes the temperature dependence of the apparent dielectric relaxation, and the activation energy of $R_1$ equals to that of the observed dielectric relaxation.

A probable structure of a sample corresponding to Fig. 5(a) is illustrated in Fig. 6, where the sample consists of two phases with different thickness and conductivity. This kind of model is known as an internal barrier layer capacitor or the Maxwell-Wagner effect. For simplicity, the intrinsic dielectric constant $\varepsilon$ for both phases is assumed to be same and independent of temperature, as differences in the electrical conductivity and its temperature dependence for each phase are essential to explain the phenomenon. In order to fulfill the condition as $R_1 \ll R_2$ and $C_1 \ll C_2$, $\sigma_1(T) \gg \sigma_2(T)$ and $p \ll 1$ are assumed. For example, the phase 1 corresponds to the semiconducting grain and the phase 2 to the insulating thin grain boundary in ceramic samples, however, the geometrical pattern of the phase 2 in the sample is not a matter here. It should be noted that the phases are not $BaTiO_3$ and $LaCrO_3$ for BLTC samples, as they are $(Ba,La)(Ti,Cr)O_3$ solid solutions.

Explicit expressions of the complex capacitance for Fig. 5(a) are given as

$$C'(\omega) = \frac{\tau_1 R_1 + \tau_2 R_2 + \omega^2 \tau_1 \tau_2 (\tau_1 R_2 + \tau_2 R_1)}{(R_1 + R_2)^2 + \omega^2 (\tau_1 R_2 + \tau_2 R_1)^2},$$

$$C''(\omega) = \frac{R_1 + R_2 + \omega^2 (\tau_1^2 R_2 + \tau_2^2 R_1)}{\omega (R_1 + R_2)^2 + \omega^3 (\tau_1 R_2 + \tau_2 R_1)^2},$$

$$\tau_1 = R_1 C_1, \ \tau_2 = R_2 C_2.$$

(3)

The circuit elements in Eq. (3) are written by the model parameters as Eq. (4).

$$C' = \varepsilon' C_0, \ C'' = \varepsilon'' C_0, \ C_0 = \varepsilon_0 S / d,$$

$$C_1 = \frac{\varepsilon C_0}{1-p}, \ C_2 = \frac{\varepsilon C_0}{p},$$

$$R_1(T) = \frac{\varepsilon_0 (1-p)}{\sigma_1 C_0}, \ R_2(T) = \frac{\varepsilon_0 p}{\sigma_2 C_0},$$

$$\sigma_1(T) = \sigma_{10} \exp(-E_1 / k_B T),$$

$$\sigma_2(T) = \sigma_{20} \exp(-E_2 / k_B T).$$

(4)

By rewriting Eq. (3) with Eq. (4), temperature and frequency dependences of $\varepsilon'$ and $\varepsilon''$ of the model can be calculated. $C_0$ is an empty cell capacitance depends on the sample geometry, however, it cancels out by rewriting. Required parameters to describe the model are six; $\varepsilon$, $\sigma_{10}$, $\sigma_{20}$, $E_1$, $E_2$, and $p$.

Values of the parameters are determined from the experimental results as follows. The intrinsic dielectric constant $\varepsilon$ equals to $\varepsilon'$ measured at lower temperature for the observed relaxation, which equals to $\varepsilon(\infty)$. The thickness ratio $p$ equals to the ratio of $\varepsilon'$ at lower and higher temperature, $\varepsilon(\infty)/\varepsilon(0)$. The conductivity $\sigma_{10}$ and its activation energy $E_1$ for phase 1 correspond to the relaxation time $\tau_0$ and the activation energy $E$ for the observed relaxation, respectively, therefore $E_1 = E$. $\sigma_{20}$ and $E_2$ are determined from $\varepsilon''$ at higher temperature. $E_2$ equals to the activation energy of $\varepsilon''(T)$. The expressions for $\sigma_{10}$ and $\sigma_{20}$ are given as

$$\sigma_{10} = \frac{(1-p)\varepsilon_0\varepsilon}{p\tau_0}, \quad \sigma_{20} = \frac{\varepsilon_0\varepsilon''(T)p\omega}{\exp(-E_2/k_BT)}. \tag{5}$$

The parameters for BLTC-0.50 are $\varepsilon = 30$, $p = 7.5 \times 10^{-3}$, $\sigma_{10} = 15.5$ S/m, $E_1 = 153$ meV, $\sigma_{20} = 1.46$ mS/m, and $E_2 = 279$ meV. The calculated result is shown in Fig. 7. The dielectric relaxation and the increase of $\varepsilon''$ at higher temperature in Fig. 3 are reproduced by this model.

The dielectric behavior of CCTO ceramics can also be explained by this model as shown in Fig. 8. The deviation of experimental results from the calculated results in $\varepsilon''(T)$ from 70 K to 220 K for 1 Hz indicates the existence of some other factors besides $\sigma_2$ and $E_2$ for $R_2$, however, $R_2$ does not affect the apparent relaxation itself and can be modified to agree with experimental $\varepsilon''(T)$ at higher temperature. The deviation below 50 K also indicates lower activation energy for $R_1$ than $E_1$. $E_1$ of the CCTO sample was determined to be 13.9 meV below 20 K by another experiment using impedance spectroscopy.

It should be noted that the complex impedance semicircle as shown in Fig. 4(a) does not always indicate the equivalent circuit as Fig. 5(a), and the circuit as Fig. 5(e) is also possible if $R_1 \ll R_2$ and $C_1 \ll C_2$. Fig. 5(e) is an equivalent circuit for the exact Debye-type dielectric relaxation and all the elements are assumed in one phase, not in separate phases. Large $C_2$ in Fig. 5(e) means high intrinsic dielectric constant, not a thin layer, however, frequency dependence of complex impedance of Fig. 5(e) is very similar to that of Fig. 5(a). It seems that definitive evidence for the barrier layer structure in CCTO has not been found. But the thin barrier layer (the phase 2 in Fig. 6) is considered to be affected by the high dc bias voltage only in a high temperature region for the observed relaxation. Because the resistance of the bulk (the phase 1) decreases and the electric field in the barrier layer is strengthened in a high temperature region, while the conductivity can be ignored and the electric field in the sample affects uniformly in a low temperature region. This expectation has been confirmed for the BLTC-0.50 and CCTO samples. Figure 9 shows the effect of dc bias voltage for the CCTO sample. Both $R_2$ and $C_2$ decrease with the dc bias at 30°C. Thickness of the phase 2 in the CCTO sample is estimated 1.7 μm by the model, then the electric field of 6 kV/mm is applied to the phase 2 with the dc bias of 10 V, which is considered the cause of this result. The dc bias effect of BLTC and CCTO samples were not observed below 20 K.

BLTC-0.03 at 30°C can be regarded as in a low temperature region for the same kind of apparent relaxation. The increase of $\varepsilon'$ seen in Fig. 2 at higher temperature can also be regarded as the same kind of behavior, however, $\varepsilon''$ due to the dc conductivity is already high and the peak due to the relaxation does not appear.

## V. CONCLUSIONS

The observed dielectric relaxation of BLTC and CCTO ceramics has been explained by a heterogeneous structure model. Not only CCTO but also many kinds of substances show the similar behavior, and the behavior can be explained universally by the model, as all the model parameters are determined from the experimental results. The high $\varepsilon'$ peak observed in $SrTiO_3$-$SrMg_{1/3}Nb_{2/3}O_3$ solid solutions[13] also seems to be explained by the model with the consideration of temperature dependence of the intrinsic dielectric constant. The keys to the behavior are the heterogeneous structure consisting of conductive bulk and insulative thin layer, and temperature dependence of the conductivity. Further studies on the origin of spontaneous formation of the heterostructure are required. Desirable dielectric properties for applications of CCTO seem to come from the high resistivity of the thin barrier layer, however, it may have a weakness on standing a high voltage.

## ACKNOWLEDGMENTS

This study was partly supported by a grant-in-aid of MEXT and a grant-in-aid for the Research and Development of New Technology from the Association of Private Universities in Japan, and the 21st century COE program (Physics of self-organization systems with multi-components) of MEXT.

We thank Dr. Guobao Li, Peking University, for supplying BLTC, and Dr. W. Kobayashi, Waseda University, for CCTO samples. We are grateful to Prof. Y. Yamada, Prof. I. Terasaki, Prof. T. Katsufuji, Waseda University, Dr. J-M. Kiat, Dr. B. Dkhil, Ecole Centrale Paris, and Prof. B. A. Strukov, Moscow State University, for valuable discussions.

TABLE I. Measured values for the dielectric relaxation of BLTC samples.

| $x$ | $\varepsilon(\infty)$ | $\varepsilon(0)$ | $\tau_0$ /ns | $E$ /meV |
|---|---|---|---|---|
| 0.40 | 40 | 3500 | 0.444 | 251 |
| 0.50 | 30 | 4000 | 2.29 | 153 |
| 0.60 | 25 | 4500 | 1.03 | 139 |
| 0.70 | 20 | 4200 | 1.11 | 132 |

**Figure captions**

Fig. 1. The crystal structure of $Ba_{1-x}La_xTi_{1-x}Cr_xO_3$ with respect to the composition ratio $x$ at RT.

Fig. 2. Temperature dependences of $\varepsilon'$ and $\varepsilon''$ of BLTC samples with different $x$. (a) $x = 0.04$, (b) $x = 0.06$, (c) $x = 0.08$, and (d) $x = 0.10$.

Fig. 3. Temperature dependences of $\varepsilon'$ (a) and $\varepsilon''$ (b) of BLTC-0.50 with various frequencies.

Fig. 4. Dependence of complex impedance ($Z'$, $Z''$) on applying dc bias with a parameter of frequencies from 1 MHz to 10 mHz at 30°C. (a) BLTC-0.50, (b) BLTC-0.03, and (c) the magnification of (a) and (b) around $|Z| = 0$ without dc bias. Note that units in each graph are different.

Fig. 5. (a) A circuit of a series of two *R-C* parallel connections, (b) a simplified circuit at low temperature or a high frequency, (c) that in the specific range of temperature and a frequency, (d) at high temperature or a low frequency, and (e) an equivalent circuit for the Debye-type dielectric relaxation with the conductivity.

Fig. 6. Model structure corresponding to Fig. 5(a). $\varepsilon$, the intrinsic dielectric constant of phase 1 and 2, $\sigma_{10}$, $\sigma_{20}$, the electrical conductivity for each phase, $E_1$, $E_2$, the activation energy of the conductivity for each phase, and $p$, the thickness ratio of phase 2 to the whole sample.

Fig. 7. Temperature dependence of calculated $\varepsilon'$ (a) and $\varepsilon''$ (b) for BLTC-0.50 for various frequencies using parameters determined by the present experiment.

Fig. 8. Temperature dependence of $\varepsilon'$ (a) and $\varepsilon''$ (b) of CCTO ceramic. Symbols are experimental results and lines are calculated ones with $\varepsilon = 100$, $p = 2.22 \times 10^{-3}$, $\sigma_{10} = 21.0$ S/m, $E_1 = 71.1$ meV, $\sigma_{20} = 76.0$ S/m, and $E_2 = 499$ meV.

Fig. 9. $Z'$-$Z''$ plot (a) and $C'$-$C''$ plot (b) of CCTO ceramic under dc bias with frequencies from 1 MHz to 100 Hz at 30°C.

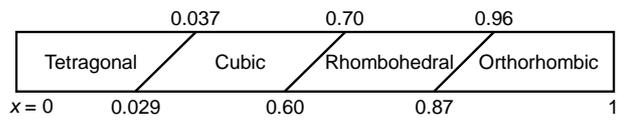

Fig. 1. M. Fukunaga and Y. Uesu.

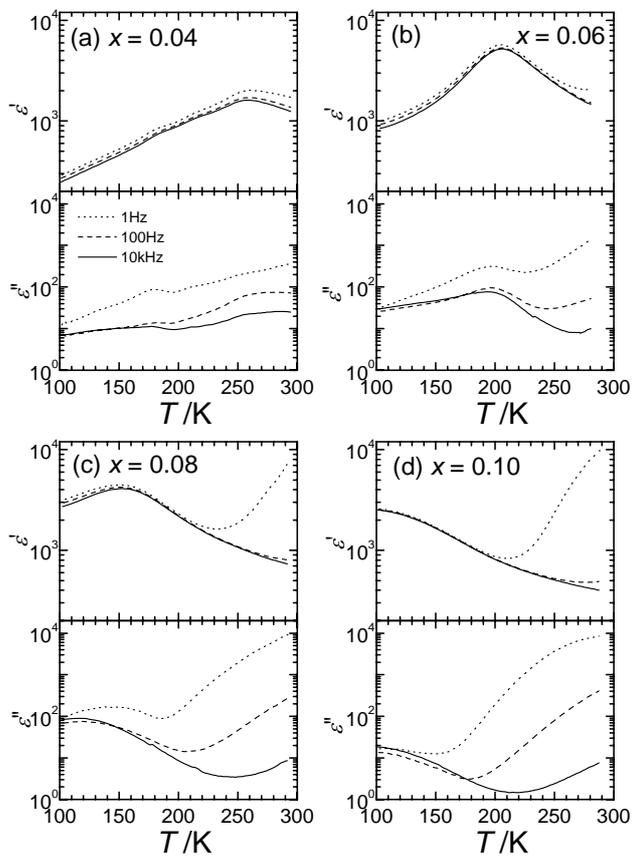

Fig. 2.  M. Fukunaga and Y. Uesu.

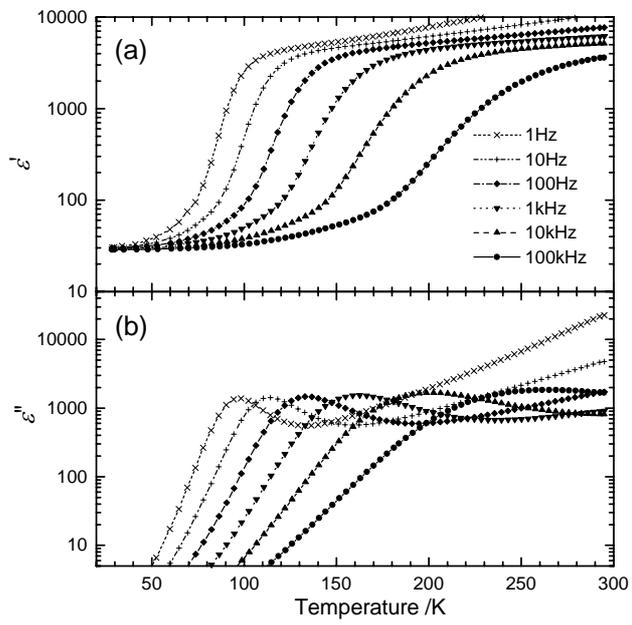

Fig. 3. M. Fukunaga and Y. Uesu.

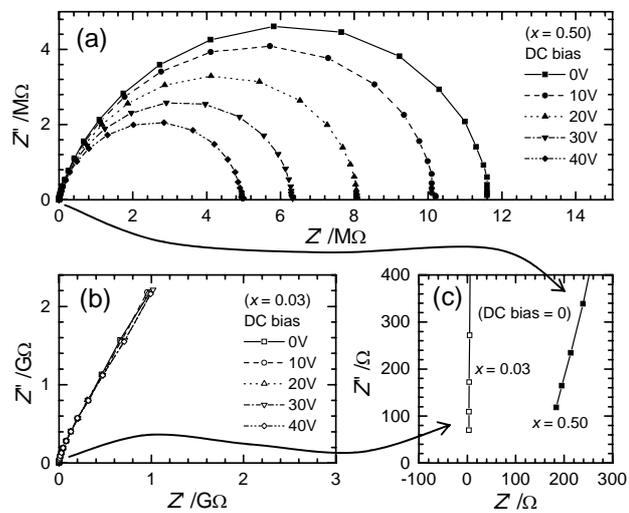

Fig. 4. M. Fukunaga and Y. Uesu.

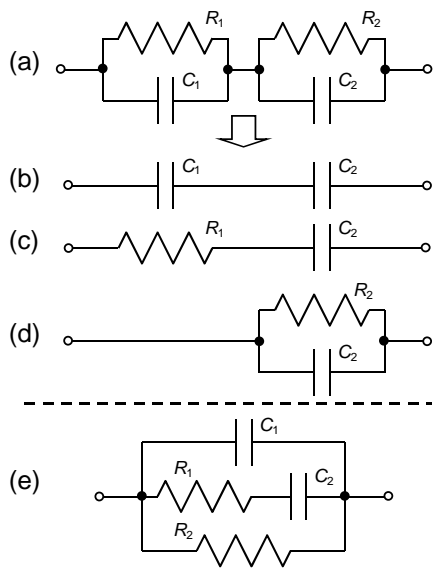

Fig. 5. M. Fukunaga and Y. Uesu.

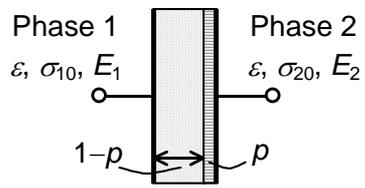

Fig. 6.  M. Fukunaga and Y. Uesu.

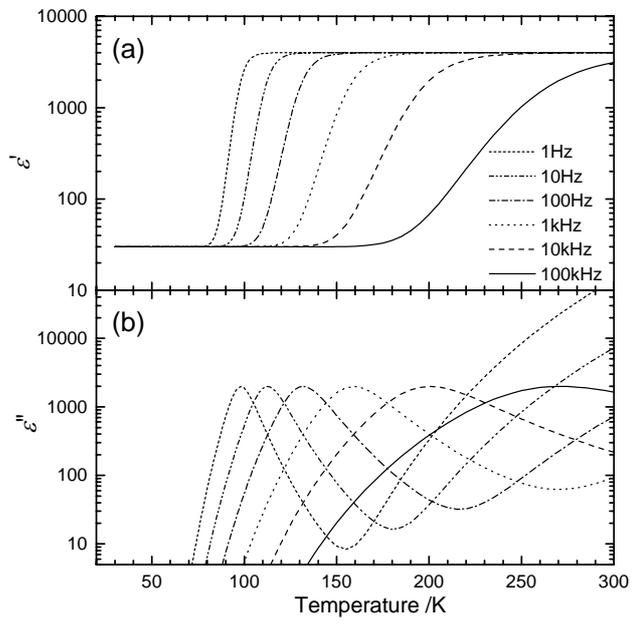

Fig. 7. M. Fukunaga and Y. Uesu.

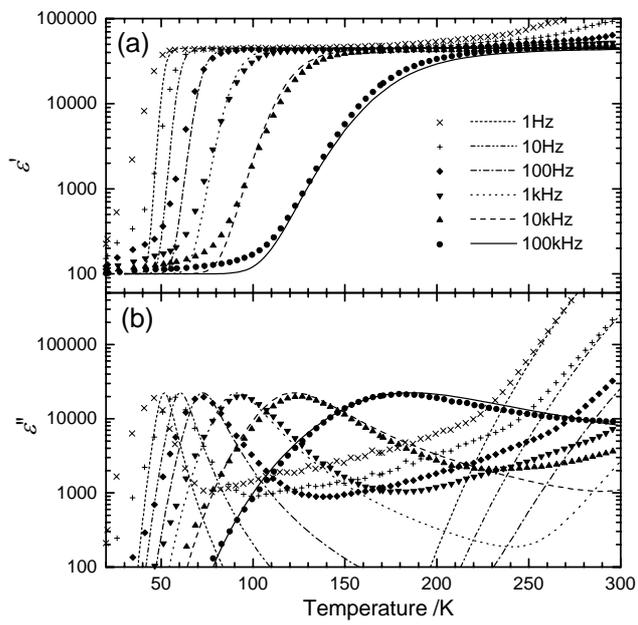

Fig. 8. M. Fukunaga and Y. Uesu.

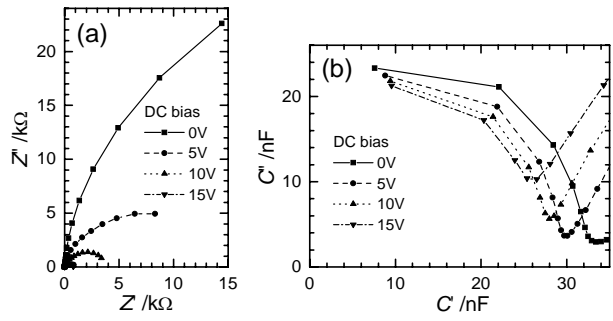

Fig. 9. M. Fukunaga and Y. Uesu.